\documentclass[twocolumn,showpacs,preprintnumbers,amsmath,amssymb,showkeys]{revtex4}

\usepackage{graphicx}
\usepackage{dcolumn}
\usepackage{bm}

\begin{document}

\preprint{}

\title{Spin-to-orbital angular momentum conversion in spin Hall effect of light}
\author{Hailu Luo}
\author{Shuangchun Wen}\email{scwen@hnu.cn}
\author{Weixing Shu}
\author{Dianyuan Fan}
\affiliation{Key Laboratory for Micro/Nano Opto-Electronic Devices
of Ministry of Education, College of Information Science and
Engineering, Hunan University, Changsha 410082, People's Republic of
China}
\date{\today}

\begin{abstract}
From the viewpoint of classical electrodynamics, we identify the
role of spin-to-orbital angular momentum conversion in spin Hall
effect (SHE) of light. We introduce a distinct separation between
spin and orbital angular momenta to clarify the spin-orbital
interaction in conventional beam refraction. We demonstrate that the
refractive index gradient can enhance or suppress the
spin-to-orbital angular momentum conversion, and thus can control
the SHE of light. We suggest that the metamaterial whose refractive
index can be tailored arbitrarily may become a good candidate for
amplifying or eliminating the SHE of light, and by properly
facilitating the spin-to-orbital angular momentum conversion the SHE
may be enhanced dramatically. The transverse spatial shifts governed
by the spin-to-orbital angular momentum conversion, provide us a
clear physical picture to clarify the role of refractive index
gradient in the SHE of light. These findings provide a pathway for
modulating the SHE of light and can be extrapolated to other
physical systems.
\end{abstract}


\maketitle

\section{Introduction}\label{SecI}
The development of spin photonics has taken an important step
forwards due to the recently experimental verifications of the spin
Hall effect (SHE) of light~\cite{Hosten2008,Bliokh2008}.  The SHE is
a transport phenomenon, in which an applied field on the spin
particles leads to a spin-dependent shift perpendicular to the
electric field~\cite{Murakami2003,Sinova2004,Wunderlich2005}. The
SHE of light can be regarded as a direct optical analogy in which
the spin electrons and electric potential are replaced by spin
photons and refractive index gradient. The SHE of light sometimes
referred to as the Fedorov--Imbert effect, was predicted
theoretically by Fedorov~\cite{Fedorov1965}, and was experimentally
confirmed by Imbert~\cite{Imbert1972}. The theory was extended in a
formulation which shows that the transverse spatial separation of
the left and right circularly polarized components on oblique
incidence directly from the total angular momentum
conservation~\cite{Onoda2004,Bliokh2006a}. More recently, the
interesting effect has also been observed in scattering from
dielectric spheres~\cite{Haefner2009}, on the direction making an
angle with the propagation axis~\cite{Aiello2009a}, and in silicon
via free-carrier absorption~\cite{Menard2010}.

The polarization-dependent transverse shift in the SHE of light is
generally believed as a result of an effective spin-orbital
interaction, which describes the mutual influence of the spin
(polarization) and trajectory of the light beam~\cite{Bliokh2008}.
There are two characteristics of the spin-orbit interaction of
photons: The first one is the influence of the trajectory upon
polarization~\cite{Chiao1986,Tomita1986}; The second one is the
reciprocal influence of the polarization upon the
trajectory~\cite{Dooghin1992,Liberman1992}. These mechanisms have
well been understood in gradient refractive index
media~\cite{Chiao1986,Tomita1986,Dooghin1992,Liberman1992}. In
addition, the spin-orbital interaction in both inhomogeneous
anisotropic media~\cite{Marrucci2006} and in tightly focused
beams~\cite{Zhao2007} can be explained by spin-to-orbital angular
momentum conversion. However, the physical picture of spin-orbit
interaction in conventional beam refraction has not yet been fully
examined. For example, the relation between refractive index
gradient and spin-to-orbital angular momentum conversion is unclear.
Whether spin-to-orbital angular momentum conversion can be enhanced
(or suppressed) by increasing (or decreasing) the refractive index
gradient? Thus, the aim of this paper is to reveal the
spin-to-orbital angular momentum conversion in the SHE of light from
the viewpoint of classical electrodynamics.

The paper is organized as follows. First, we establish a three
dimension beam propagation model for describing the SHE of light in
conventional beam refraction. Our result shows that the refractive
index gradient can enhance or suppress the transverse spatial shifts
in SHE of light. We suggest that the metamaterial whose refractive
index can be tailored arbitrarily is a good candidate to modulate
the SHE of light. Next, we attempt to obtain a clear physical
picture of spin-to-orbital angular momentum conversion in the SHE of
light. Within the paraxial approximation, a distinct separation
between spin and orbital angular momenta is introduced. We find that
the SHE of light may be dramatically enhanced by facilitating the
spin-to-orbital angular momentum conversion. Finally, we want to
explore what role refractive-index gradient plays in the
spin-to-orbital angular momentum conversion. We demonstrate that the
refractive index gradient can enhance or suppress the
spin-to-orbital angular momentum conversion, and thus can control
the SHE of light.

\section{Three dimension beam propagation model}\label{SecII}
\begin{figure}
\includegraphics[width=8cm]{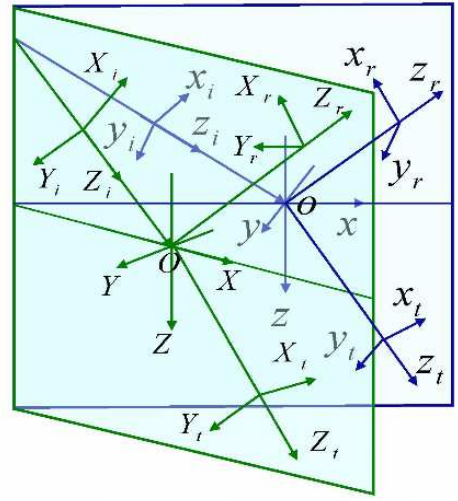}
\caption{\label{Fig1} (color online) Schematic illustrating the
reflection and refraction of central and local wave vectors at an
air-glass interface in Cartesian coordinate system. $x_ay_az_a$ are
are reference frames for central wave vector and $X_aY_aZ_a$ are
local reference frames for an arbitrary wave vector. $oxyz$ and
$OXYZ$ are the interface reference frames for central and local wave
vectors, respectively.}
\end{figure}

To reveal the spin-to-orbital angular momentum conversion, we need
to establish a three dimension beam propagation model for describing
the SHE of light. Figure~\ref{Fig1} illustrates the beam reflection
and refraction in Cartesian coordinate system. The $z$ axis of the
laboratory Cartesian frame ($x,y,z$) is normal to the air-glass
interface locating at $z=0$. We use the coordinate frames
($x_a,y_a,z_a$) for central wave vector, where $a=i,r,t$ denotes
incident, reflection, and transmission, respectively. In addition,
we must introduce local Cartesian frames ($X_a,Y_a,Z_a$) to describe
an arbitrary wave vector. The electric field of the $a$th beam can
be solved by employing the Fourier
transformations~\cite{Goodman1996}. The complex amplitude for the
$a$th beam can be conveniently expressed as
\begin{eqnarray}
\mathbf{E}_a(x_a,y_a,z_a )&=&\int d k_{ax}dk_{ay}
\tilde{\mathbf{E}_a}(k_{ax},k_{ay})\nonumber\\
&&\times\exp [i(k_{ax}x_a+k_{ay}y_a+ k_{az} z_a)],\label{noapr}
\end{eqnarray}
where $k_{az}=\sqrt{k_a^2-(k_{ax}^2+k_{ay}^2)}$ and
$\tilde{\mathbf{E}_a}(k_{ax},k_{ay})$ is the angular spectrum. The
approximate paraxial expression for the field in Eq.~(\ref{noapr})
can be obtained by the expansion of the square root of $k_{az}$ to
the first order~\cite{Lax1975}, which yields
\begin{eqnarray}
\mathbf{E}_a&=&\exp(i k_a z_a) \int dk_{ax}dk_{ay}
\tilde{\mathbf{E}_a}(k_{ax},k_{ay})\nonumber\\
&&\times\exp \left[i\left(k_{ax}x_a+k_{ay}y_a-\frac{k_{ax}^2+k_{a
y}^2}{2 k_a}z_a\right)\right]\label{apr}.
\end{eqnarray}
In general, an arbitrary linear polarization can be decomposed into
horizontal and vertical components. In the spin basis set, the
angular spectrum can been written as:
\begin{equation}
\tilde{E}_i^H=\frac{1}{\sqrt{2}}(\tilde{\mathbf{E}}_{i+}+\tilde{\mathbf{E}}_{i-})\label{SBH},
\end{equation}
\begin{equation}
\tilde{E}_i^V=\frac{1}{\sqrt{2}}i(\tilde{\mathbf{E}}_{i-}-\tilde{\mathbf{E}}_{i+})\label{SBV}.
\end{equation}
Here, $H$ and $V$ represent horizontal and vertical polarizations,
respectively. The positive and negative signs denote the left and
right components, respectively~\cite{Beth1936}. The monochromatic
Gaussian beam can be formulated as a localized wave packet whose
spectrum arbitrarily narrows, whose angular spectrum can be written
as
\begin{equation}
\tilde{\mathbf{E}}_{i\pm}=(\mathbf{e}_{ix} \pm
i\mathbf{e}_{iy})\frac{w_0}{\sqrt{2\pi}}\exp\left[-\frac{w_0^2(k_{ix}^2+k_{iy}^2)}{4}\right]\label{asi},
\end{equation}
where $w_0$ is the beam waist. After the angular spectrum is known,
we can obtain the field characteristics for the $a$th beam.

To accurately describe the SHE of light, it is need to determine the
reflection and transmission of arbitrary wave-vector components.
From the central frame $x_iy_iz_i$ to the local frame $X_iY_iZ_i$,
the following three steps should be carried out. First, we transform
the electric field from the reference frame $x_iy_iz_i$ around the
$y$ axis by the incident angle $\theta_i$ to the frame $xyz$:
$\tilde{E}_{xyz}=m_{{x_iy_iz_i}\rightarrow{xyz}}\tilde{E}_{x_iy_iz_i}$,
where
\begin{eqnarray}
m_{{x_iy_iz_i}\rightarrow{xyz}}=\left[
\begin{array}{ccc}
\cos\theta_i & 0 & -\sin\theta_i\\
0 & 1 & 0\\
\sin\theta_i & 0 & \cos\theta_i
\end{array}
\right].\label{matrixr}
\end{eqnarray}
Then, we transform the electric field from the reference frame $xyz$
around the $y$ axis by an angle $k_{iy}/k_0\sin\theta_i$ to the
frame $XYZ$, and the correspondingly matrix is given by
\begin{eqnarray}
m_{{xyz}\rightarrow{XYZ}}=\left[
\begin{array}{ccc}
1 & \frac{k_{iy} }{k_0\sin\theta_i} & 0\\
-\frac{k_{iy}}{k_0\sin\theta_i} & 1 & 0\\
0 & 0 & 1
\end{array}
\right].\label{matrixr}
\end{eqnarray}
Finally, we transform the electric field from the reference frame
$XYZ$ around the $y$ axis by an angle $-\theta_i$ to the frame
$X_iY_iZ_i$, and the rotation matrix can be written as
\begin{eqnarray}
m_{{XYZ}\rightarrow{X_iY_iZ_i}}=\left[
\begin{array}{ccc}
\cos\theta_i & 0 & \sin\theta_i\\
0 & 1 & 0\\
-\sin\theta_i & 0 & \cos\theta_i
\end{array}
\right].\label{matrixr}
\end{eqnarray}
The rotation matrix of the reference frame $x_iy_iz_i$ around the
$y$ axis by an angle
$M_{{x_iy_iz_i}\rightarrow{X_iY_iZ_i}}=m_{{XYZ}\rightarrow{X_iY_iZ_i}}
m_{{xyz}\rightarrow{XYZ}}m_{{x_iy_iz_i}\rightarrow{xyz}}$, and we
have
\begin{eqnarray}
M_{{x_iy_iz_i}\rightarrow{X_iY_iZ_i}}=\left[
\begin{array}{cc}
1 &\frac{k_{ry} \cot\theta_i}{k_0} \\
-\frac{k_{ry} \cot\theta_i}{k_0} & 1
\end{array}
\right].\label{matrixr}
\end{eqnarray}
For an arbitrary wave vector, the reflected field is determined by
$\tilde{E}_{X_rY_rZ_r}=r_{p,s}\tilde{E}_{X_iY_iZ_i}$, where $r_p$
and $r_s$ are the Fresnel reflection coefficients. The reflected
field should be transformed from $X_rY_rZ_r$ to $x_ry_rz_r$.
Following the similar procedure, the reflected field can be obtained
by carrying out three steps of transformation:
$\tilde{E}_{x_ry_rz_r}=m_{{X_rY_rZ_r}\rightarrow{x_ry_rz_r}}\tilde{E}_{X_rY_rZ_r}$
where
\begin{eqnarray}
M_{{X_rY_rZ_r}\rightarrow{x_ry_rz_r}}=\left[
\begin{array}{cc}
1 &\frac{k_{ry} \cot\theta_i}{k_0} \\
-\frac{k_{ry} \cot\theta_i}{k_0} & 1
\end{array}
\right].\label{matrixr}
\end{eqnarray}
Here, only the two dimension rotation matrices is taken into
account, since the longitudinal component of electric field can be
obtained from the divergence equation
$\tilde{E}_{az}k_{az}=-(\tilde{E}_{ax}k_{ax}+\tilde{E}_{ay}k_{ay})$.
The reflection matrix can be written as
\begin{eqnarray}
M_R=M_{{X_rY_rZ_r}\rightarrow{x_ry_rz_r}}\left[
\begin{array}{cc}
r_p &0 \\
0 & r_s
\end{array}
\right]M_{{x_iy_iz_i}\rightarrow{X_iY_iZ_i}}.\label{matrixrI}
\end{eqnarray}
The reflected angular spectrum is related to the boundary
distribution of the electric field by means of the relation
$\tilde{E}_r(k_{rx},k_{ry})=M_R\tilde{E}_i(k_{ix},k_{iy})$, and we
have
\begin{eqnarray}
\left[\begin{array}{cc}
\tilde{E}_r^H\\
\tilde{E}_r^V
\end{array}\right]
=\left[
\begin{array}{cc}
r_p&\frac{k_{ry} (r_p+r_s) \cot\theta_i}{k_0} \\
-\frac{k_{ry} (r_p+r_s)\cot\theta_i}{k_0} & r_s
\end{array}
\right]\left[\begin{array}{cc}
\tilde{E}_i^H\\
\tilde{E}_i^V
\end{array}\right],\label{matrixrII}
\end{eqnarray}
It is well known that $r_p$ and $r_s$ can be expanded as a power
series that can be truncated to an appropriate order $N$. By making
use of Taylor series expansion based on the arbitrary angular
spectrum component, $r_p$ and $r_s$ can be expanded as a polynomial
of $k_{ix}$:
\begin{eqnarray}
r_{p,s}(k_{ix})&=&r_{p,s}(k_{ix}=0)+k_{ix}\left[\frac{\partial
r_{p,s}(k_{ix})}{\partial
k_{ix}}\right]_{k_{ix}=0}\nonumber\\
&&+\sum_{j=2}^{N}\frac{k_{ix}^N}{j!}\left[\frac{\partial^j
r_{p,s}(k_{ix})}{\partial k_{ix}^j}\right]_{k_{ix}=0}\label{LMD}.
\end{eqnarray}
Our analysis is confined to the zero order to obtain a sufficiently
good approximation. From the boundary condition, we obtain
$k_{rx}=-k_{ix} $ and $k_{ry}= k_{iy}$. In fact, after the incident
angular spectrum is known, Eq.~(\ref{apr}) together with
Eqs.~(\ref{asi}) and (\ref{matrixrII}) provides the paraxial
expression of the reflected field:
\begin{eqnarray}
\mathbf{E}_{r\pm}^H&=&\frac{r_p}{\sqrt{\pi
}w_0}\frac{z_R}{z_R+iz_r}\exp(ik_rz_r)\nonumber\\
&&\times\exp\left[-\frac{k_0}{2}\frac{x_{r}^2+(y_{r}\pm\delta_r^H)^2}{z_R+iz_r}\right]\label{HPR}.
\end{eqnarray}
\begin{eqnarray}
\mathbf{E}_{r\pm}^V&=&\frac{\pm i r_s}{\sqrt{\pi
}w_0}\frac{z_R}{z_R+iz_r}\exp(ik_rz_r)\nonumber\\
&&\times\exp\left[-\frac{k_0}{2}\frac{x_{r}^2+(y_{r}\pm\delta_r^V)^2}{z_R+iz_r}\right]\label{VPR}.
\end{eqnarray}
where $\delta_r^H=(1+r_s/r_p)\cot\theta_i/k$ and
$\delta_r^V=(1+r_s/r_p)\cot\theta_i/k$. Note that the above
expression of reflected field coincides with the early
results~\cite{Bliokh2007,Aiello2008,Menzel2008,Luo2009} with
different methods.

We proceed to consider the transmitted field. Following the similar
procedure, we obtain the transform matrix from $X_tY_tZ_t$ to
$x_ty_tz_t$ as
\begin{eqnarray}
M_{{X_tY_tZ_t}\rightarrow{x_ty_tz_t}}=\left[
\begin{array}{cc}
1 &\frac{k_{ty} \cos\theta_t}{k_0 \sin\theta_i} \\
-\frac{k_{ty} \cos\theta_t}{k_0 \sin\theta_i} & 1
\end{array}
\right],\label{matrixr}
\end{eqnarray}
where $\theta_t$ is the transmitted angle. For an arbitrary wave
vector, the transmitted field is determined by
$\tilde{E}_{X_tY_tZ_t}=t_{p,s}\tilde{E}_{X_iY_iZ_i}$, where $t_p$
and $t_s$ are the Fresnel transmission coefficients. Hence, the
transmitted field should be transformed from $X_rY_rZ_r$ to
$x_ty_tz_t$, and the transmission matrix can be written as
\begin{eqnarray}
M_T=M_{{X_tY_tZ_t}\rightarrow{x_ty_tz_t}}\left[
\begin{array}{cc}
t_p &0 \\
0 & t_s
\end{array}
\right]M_{{x_iy_iz_i}\rightarrow{X_iY_iZ_i}}.\label{matrixtI}
\end{eqnarray}
The transmitted angular spectrum is related to the boundary
distribution of the electric field by means of the relation
$\tilde{E}_t(k_{tx},k_{ty})=M_T\tilde{E}_i(k_{ix},k_{iy})$, and can
be written as
\begin{eqnarray}
\left[\begin{array}{cc}
\tilde{E}_t^H\\
\tilde{E}_t^V
\end{array}\right]
=\left[
\begin{array}{cc}
t_p&\frac{k_{ty} (t_p-\eta t_s) \cot\theta_i}{k_0} \\
\frac{k_{ty} (\eta t_p-t_s)\cot\theta_i}{k_0} & t_s
\end{array}
\right]\left[\begin{array}{cc}
\tilde{E}_i^H\\
\tilde{E}_i^V
\end{array}\right],\label{matrixtII}
\end{eqnarray}
where $\eta=\cos\theta_t/\cos\theta_i$. Based on the Taylor series
expansion, the Fresnel transmission coefficients $t_p$ and $t_s$ can
be written as
\begin{eqnarray}
t_{p,s}(k_{ix})&=&t_{p,s}(k_{ix}=0)+k_{ix}\left[\frac{\partial
t_{p,s}(k_{ix})}{\partial
k_x}\right]_{k_{ix}=0}\nonumber\\
&&+\sum_{j=2}^{N}\frac{k_{ix}^N}{j!}\left[\frac{\partial^j
t_{p,s}(k_{ix})}{\partial k_{ix}^j}\right]_{k_{ix}=0}\label{TPS}.
\end{eqnarray}
Note that the Fresnel coefficients are real in the regime of partial
reflection and transmission, $\sin\theta_i<n$. From the Snell's law
under the paraxial approximation, we obtain $k_{tx}=k_{ix}/\eta $
and $k_{ty}= k_{iy}$. Substituting Eqs.~(\ref{asi}) and
(\ref{matrixtII}) into Eq.~(\ref{apr}), we obtain the transmitted
field:
\begin{eqnarray}
\mathbf{E}_{t\pm}^H&=&\frac{t_p}{\sqrt{\pi} w_0}\frac{ z_{Ry} \exp
(ik_tz_t)}{\sqrt{(z_{Rx}+iz_t)(z_{Ry}+iz_t)}}
\nonumber\\&&\times\exp\left[-\frac{n k_0}{2}\left(\frac{x_{t}^2}{
z_{Rx}+iz_t}+\frac{(y_{t}\pm\delta_t^H)^2}{z_{Ry}+iz_t}\right)\right]\label{HPT}.
\end{eqnarray}
\begin{eqnarray}
\mathbf{E}_{t\pm}^V&=&\frac{\pm i t_s}{\sqrt{\pi} w_0}\frac{z_{Ry}
\exp (ik_tz_t)}{\sqrt{(z_{Rx}+iz_t)(z_{Ry}+iz_t)}}
\nonumber\\&&\times\exp\left[-\frac{n k_0}{2}\left(\frac{x_{t}^2}{
z_{Rx}+iz_t}+\frac{(y_{t}\pm\delta_t^V)^2}{z_{Ry}+iz_t}\right)\right]\label{VPT}.
\end{eqnarray}
Here, $\delta_t^H=(1-\eta t_s/t_p)\cot\theta_i/k$ and
$\delta_t^V=(1-\eta t_p/t_s)\cot\theta_i/k$. The interesting point
we want to stress is that there are two different Rayleigh lengths,
$z_{Rx}=n\eta^2k_0w_0^2/2$ and $z_{Ry}=n k_0w_0^2/2$, characterizing
the spreading of the beam in the direction of $x$ and $y$ axes,
respectively. Up to now, we have established a general propagation
model to describe the reflected and transmitted fields.

\section{Role of the refractive index gradient}\label{SecII}
It is well known that the SHE of light manifests itself as
polarization-dependent transverse shifts in the process of
reflection and refraction. To reveal the SHE of light, we now
determine the transverse spatial shifts of field centroid. The
time-averaged linear momentum density associated with the
electromagnetic field can be shown to be~\cite{Jackson1999}
\begin{equation}
\mathbf{p}_{a}(\mathbf{r})=
\frac{1}{2c^2}\mathrm{Re}[\mathbf{E}_{a}(\mathbf{r})
\times\mathbf{H}_{a}^\ast(\mathbf{r})]\label{LMD},
\end{equation}
where the magnetic field can be obtained by
$\mathbf{H}_{a}=-ik_{a}^{-1} \nabla\times\mathbf{E}_{a}$. The
intensity distribution of electromagnetic fields is closely linked
to the longitudinal momentum currents
$I(x_a,y_a,z_a)\propto\mathbf{p}_a\cdot \mathbf{e}_{az}$.

\begin{figure}
\includegraphics[width=8cm]{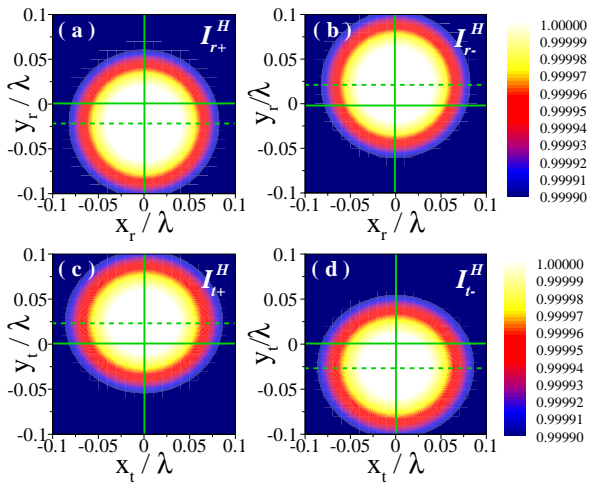}
\caption{\label{Fig2} (color online) The SHE of light manifests
itself as polarization-dependent transverse shifts of field
centroid. [(a), (b)] Intensity distribution of the reflected field
for left and right circularly polarized components, respectively.
[(c), (d)] Intensity distribution of the transmitted field for left
and right circularly polarized components, respectively. The
refractive index of the glass is $n=1.515$ and the incident angle is
chosen as $\theta_i=\pi/6$. The beam waist is chosen as
$w_0=10\lambda$. The intensity distributions in the plane $z_a=0$
are plotted in normalized units.}
\end{figure}

Figure~\ref{Fig2} shows the transverse spatial shifts of beam
centroid in an air-glass interface. In the case of reflection, the
left circularly polarized component exhibits a negative shift
[Fig.~\ref{Fig2}(a)]. For the right circular polarization, however,
presents a positive shift [Fig.~\ref{Fig2}(b)]. In the case of
transmission, the left circularly polarized component exhibits a
positive transverse shift [Fig.~\ref{Fig2}(c)]. For the right
circularly polarized component, however, presents a negative
transverse shift [Fig.~\ref{Fig2}(d)]. This interesting phenomenon
gives a clear evidence of polarization-dependent splitting of field
intensity in the SHE of light. The transverse shifts are
polarization-dependent, and thus can be regarded as the influence of
the polarization upon trajectory. In the air-glass interface, the
transverse shifts are just a few tens of nanometers, which can be
observed via weak measurements~\cite{Hosten2008,Krowne2009}.
However, how to amplify this tiny effect is still an open problem.

As we know that the refractive index gradient acts as the electric
potential gradient in the electronic systems. Now a question
naturally arises: Whether can the refractive index gradient enhance
the transverse shift in the SHE of light? To answer this question we
need to obtain a relation between the transverse shift and the
refractive index gradient. At any given plane $z_a=\text{const.}$,
the transverse spatial shift of beam centroid compared to the
geometrical-optics prediction is given by
\begin{equation}
\Delta y_{a}= \frac{\int \int y_a I(x_a,y_a,z_a) \text{d}x_a
\text{d}y_a}{\int \int I(x_a,y_a,z_a) \text{d}x_a
\text{d}y_a}.\label{centroid}
\end{equation}
Note that the transverse spatial shift is $z_a$-independent, while
the transverse angular shift can be regarded as a small shift
inclining from the $z_a$ axis. In addition, the field also
experience a longitudinal spatial
shift~\cite{Goos1974,Merano2007,Lofler2010} and a longitudinal
angular shift~\cite{Merano2009,Aiello2009b}. Note that the angle
shift means that the Snell's law cannot accurately describe the beam
refraction phenomenon~\cite{Duval2006}.

We first consider the transverse spatial shift of the reflected
field. After substituting the reflected field Eqs.~(\ref{HPR}) and
(\ref{VPR}) into Eq.~(\ref{centroid}), we obtain the transverse
spatial shifts as
\begin{equation}
\Delta y_{r\pm}^H =\mp\frac{\lambda}{2\pi} (1+r_s/r_p)\cot
\theta_i\label{TSRH},
\end{equation}
\begin{equation}
\Delta y_{r\pm}^V =\mp\frac{\lambda}{2\pi} (1+r_p/r_s)\cot
\theta_i\label{TSRV}.
\end{equation}
For an arbitrary linearly polarized beam, the transverse spatial
shift of the reflected field is given by
\begin{equation}
\Delta y_{r\pm}=\cos\gamma_r^2\Delta y_{r\pm}^H+\sin\gamma_r^2\Delta
y_{r\pm}^V\label{TSRHV},
\end{equation}
where $\gamma_r$ is the reflected polarization angle. In the frame
of classical electrodynamics, the reflection polarization angle is
determined by:
\begin{equation}
\cos\gamma_r=\frac{\cos\gamma_i
r_p}{\sqrt{\cos^2\gamma_i^2r_p^2+\sin^2\gamma_i^2r_s^2}},
\end{equation}
\begin{equation}
\sin\gamma_r=\frac{\sin\gamma_i
r_s}{\sqrt{\cos^2\gamma_i^2r_p^2+\sin^2\gamma_i^2r_s^2}}.
\end{equation}
Here, $\gamma_i$ is the incident polarization angle. Under the limit
of ultra-high refractive index gradient, the transverse spatial
shifts can be written as $\Delta y_{r\pm}=0$. This simple result
means that the recent advent of a new class of metamaterial with
ultra-high refractive index~\cite{Shin2009} is a possible candidate
to eliminate the transverse spatial shift in the reflected field.

\begin{figure}
\includegraphics[width=8cm]{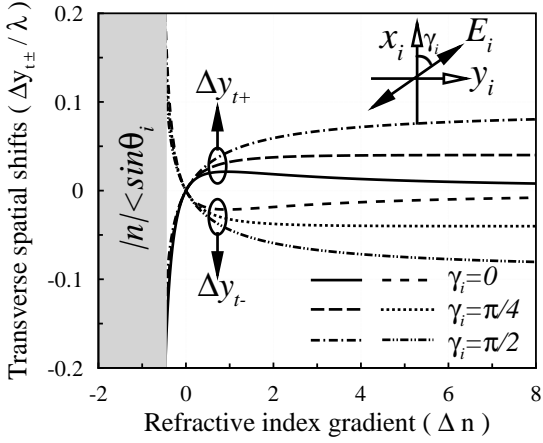}
\caption{\label{Fig3} (a) Schematically shows the SHE of light
manifests itself as spin-dependent splitting. The transverse spatial
shifts of transmitted field $\Delta{y_{t\pm}}/\lambda$ versus the
thickness of air gap $d_2$. The incident polarization angles are
chooser as (b) $\theta_i=0$ , (c)  $\theta_i=\pi/4$, and
$\theta_i=\pi/2$. Other parameters are chosen to be the same as in
Fig.~\ref{Fig2}.}
\end{figure}

We next consider the transverse spatial shifts of the transmitted
field. After substituting the transmitted field Eqs.~(\ref{HPT}) and
(\ref{VPT}) into Eq.~(\ref{centroid}), we have
\begin{equation}
\Delta y_{t\pm}^H =\mp\frac{\lambda}{2\pi} (1-\eta t_s/t_p)\cot
\theta_i\label{TSRH},
\end{equation}
\begin{equation}
\Delta y_{t\pm}^V =\mp\frac{\lambda}{2\pi} (1-\eta t_p/t_s)\cot
\theta_i\label{TSRV}.
\end{equation}
For an arbitrary linearly polarized wave-packet, the transverse
spatial shifts of the transmitted field are given by
\begin{equation}
\Delta y_{t\pm}=\cos\gamma_t^2\Delta y_{t\pm}^H+\sin\gamma_t^2\Delta y_{t\pm}^V\label{TSRHV},
\end{equation}
where the transmission polarization angle $\gamma_t$ is determined
by
\begin{equation}
\cos\gamma_t=\frac{\cos\gamma_i
t_p}{\sqrt{\cos^2\gamma_i^2t_p^2+\sin^2\gamma_i^2t_s^2}}\label{COST},
\end{equation}
\begin{equation}
\sin\gamma_t=\frac{\sin\gamma_i
t_s}{\sqrt{\cos^2\gamma_i^2t_p^2+\sin^2\gamma_i^2t_s^2}}\label{SINT}.
\end{equation}
Note that the above expression coincides well with the early
results~\cite{Hosten2008} with the quantum method. Our scheme shows
that the SHE of light can be explained from the classic
electrodynamics. Under the limit of ultra-high refractive index
gradient, the transmitted field tends to reach a saturation value:
\begin{equation}
\Delta y_{t\pm}=\pm\frac{\lambda\sin\theta_i\cos^2\gamma_i^2}
{2\pi(\cos^2\gamma_i^2+\cos^2\theta_i^2\sin^2\gamma_i^2)}\label{TSTL}.
\end{equation}
Up to now, we have described the polarization-dependent splitting in
the SHE of light from the viewpoint of pure classical
electrodynamics.

We proceed to examine the role of refractive index gradient (i.e.,
$\Delta{n}=|n|-1$) in the SHE of light. The normalized transverse
spatial shifts of transmitted field for various refractive index
gradients as shown in Fig.~\ref{Fig3}. We first consider the wave
packet incident from air to a low-refractive-index medium
($\Delta{n}<0$). For the left circularly polarized component, the
field centroid exhibits a negative transverse shift. For the right
circularly polarized component, the beam centroid also presents a
transverse spatial shift, but in an opposite direction. We next
consider the wave packet incident from air to a
high-refractive-index medium ($\Delta{n}>0$). For a certain
polarized component, we find that the SHE of light is reversed when
the refractive index gradient is inverted. It is clearly shown that
the transverse spatial shifts increases with the increase of the
refractive index gradient $\Delta n$. Within the regime of low
refractive index, the transverse spatial shifts are enhanced sharply
with the increase of $\Delta n$. When the refractive index
$|n|<\sin\theta_i$, the beam is totally reflected. While in the
regime of high refractive index, the transverse spatial shifts tend
to reach a saturation value. In addition, the transverse spatial
shifts are also sensitive to the incident polarization angle as
clearly shown in the figure. Hence, we can enhance or suppress the
transverse spatial shifts in SHE of light by modulating the
refractive index gradient and incident polarization angle.

\begin{figure}
\includegraphics[width=8cm]{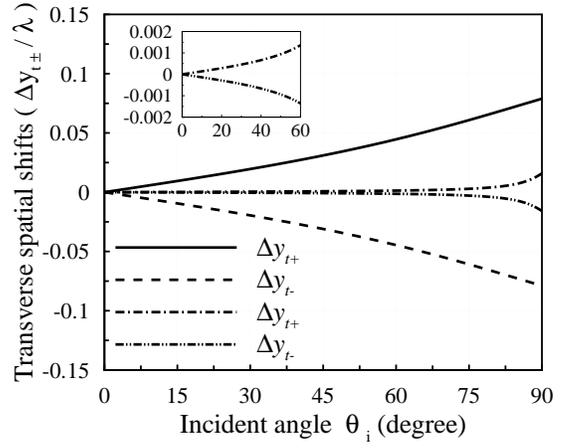}
\caption{\label{Fig4} The normalized transverse spatial shifts of
transmitted field $\Delta y_{t\pm}/\lambda$ versus incident angles
$\theta_i$ in the metamaterial with $\varepsilon=-1.00+0.028i$ and
$\mu=-1.01+0.028i$ (dashed-dotted and dashed-doted-dotted lines). As
a comparison, the conventional medium is chosen as the glass with
$n=+1.515$ (solid and dashed lines). The incident polarization angle
choose as $\gamma_i=0$. Other parameters are chosen to be the same
as in Fig.~\ref{Fig2}. Inset: Zoomed-in view of the transverse
spatial shifts at the air-metamaterial interface. }
\end{figure}

How to attenuate the SHE of light is not of pure academic interest,
owing to the requirement of eliminating the transverse shifts, such
as reflection, refraction, and focusing in optical experiments. We
suggest that the metamaterial whose refractive index can be tailored
arbitrarily~\cite{Smith2004,Pendry2006} may become a good candidate.
In order to accurately describe how the metamaterial attenuate the
SHE of light, it is necessary to include material dispersion and
losses. Thus, a certain dispersion relation, such as the Lorentz
medium model, should be introduced. The relative constitutive
parameters are
\begin{eqnarray}
\varepsilon(\omega)&=&1-\frac{\omega_{ep}^2}{\omega^2
-\omega_{eo}^2+i\Gamma_e\omega },\\
\mu(\omega)&=&1-\frac{F\omega_{mp}^2}{\omega^2-\omega_{mo}^2+i\Gamma_m\omega
}.
\end{eqnarray}
To avoid the trouble involving in a certain value of frequency, we
assume the material parameters are
$\omega_{eo}=\omega_{mo}=\omega_o$, $\omega_{ep}^2=\omega_{mp}^2=2
\omega_o^2$, $F=1.005$, and $\Gamma_e=\Gamma_m=0.01\omega_0$.
Figure~\ref{Fig4} shows the normalized transverse spatial shifts of
transmitted field versus incident angle at the air-metamaterial
interface. For the comparison, we also plot the transverse spatial
shifts at the air-glass interface. In a large range of incident
angles, the metamaterial is a good candidate for suppressing the SHE
of light in the process of refraction. In the ideal case
($\varepsilon=-1$ and $\mu=-1$), the transverse shift can be
eliminated completely. This is another reason why a simple planar
slab provides us with a perfect lens without
aberration~\cite{Pendry2000}. As we will see in the following, the
spin-to-orbital angular momentum conversion can be used to explain
the inherent physics underlying these intriguing phenomena.

\section{Spin-to-orbital angular momentum conversion}\label{SecII}
It should be noted is that the spin-orbital interaction in both
inhomogeneous anisotropic media~\cite{Marrucci2006} and in tightly
focused beams~\cite{Zhao2007} can be explained by spin-to-orbital
angular momentum conversion. To obtain a clear physical picture of
the SHE of light, we introduce a distinct separation between spin
and orbital angular momenta to clarify the spin-orbital interaction.
The momentum current can be regarded as the combined contributions
of spin and orbital parts:
\begin{equation}
\mathbf{p}_a=\mathbf{p}_a^O+\mathbf{p}_a^S.
\end{equation}
Here, the orbital term is determined by the macroscopic energy
current with respect to an arbitrary reference point and does not
depend on the polarization. The spin term, on the other hand,
relates to the phase between orthogonal field components and is
completely determined by the state of
polarization~\cite{Bekshaev2007}. In a monochromatic optical beam,
the spin and orbital currents can be respectively written in the
form:
\begin{equation}
\mathbf{p}_a^{O}=\mathrm{Im}[\mathbf{E}_a^{\ast}\cdot(\nabla)\mathbf{E}_a],
\end{equation}
\begin{equation}
\mathbf{p}_a^{S}=\mathrm{Im}[(\mathbf{E}_a\cdot\nabla)\mathbf{E}_a^{\ast}],
\end{equation}
where
$\mathbf{E}_a^{\ast}\cdot(\nabla)\mathbf{E}_a=E_{ax}^{\ast}\nabla
E_{ax}+E_{ay}^{\ast}\nabla E_{ay}+E_{az}^{\ast}\nabla E_{az}$ is the
invariant Berry notation~\cite{Berry2009}. It has been shown that
both spin and orbital currents originate from the beam transverse
inhomogeneity and their components are directly related to the
azimuthal and radial derivatives of the beam profile parameters. The
orbital currents are mainly produced by the phase gradient, while
the spin currents are orthogonal to the intensity gradient.

\begin{figure}
\includegraphics[width=8cm]{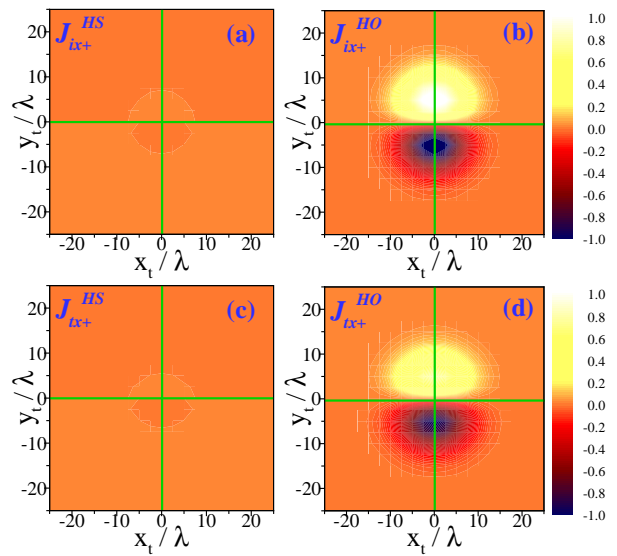}
\caption{\label{Fig5} (color online) The transverse angular momentum
density of left circularly polarized component. [(a) and (b)]
Incident fields: $j_{ix+}^{HS}$ and $j_{ix+}^{HO}$. [(c) and (d)]
Transmitted fields: $j_{tx+}^{HS}$ and $j_{tx+}^{HO}$. The cross
section is chosen as $z_t=0$ and the intensity is plotted in
normalized units. Other parameters are the same as those in
Fig.~\ref{Fig2}.}
\end{figure}

We proceed to analysis the angular momentum density for each of
individual wave packet, which can be written as
\begin{equation}
\mathbf{j}_a(\mathbf{r})=\mathbf{r}_a\times\mathbf{p}_a(\mathbf{r}_a)\label{AMD}.
\end{equation}
Within the paraxial approximation, the angular momentum can be
divided into the orbital and spin angular parts:
$\mathbf{j}_a=\mathbf{j}_a^O+\mathbf{j}_a^S$~\cite{Allen1992,Padgett1995},
it follows that
\begin{equation}
\mathbf{j}_a^O=\mathbf{r}_a\times\mathbf{p}_a^O\label{AMDO},
\end{equation}
\begin{equation}
\mathbf{j}_a^S=\mathbf{r}_a\times\mathbf{p}_a^S\label{AMDS}.
\end{equation}
It should be mentioned that this separation still hold beyond the
paraxial approximation~\cite{Barnett2002}. The transverse angular
momentum density $j_x$ can be regarded as the combined contributions
of orbital and spin parts:
\begin{equation}
j_{ax}^O=y_a p_{az}^O-z_a p_{ay}^O\label{OAMDX},
\end{equation}
\begin{equation}
j_{ax}^S=y_a p_{az}^S-z_a p_{ay}^S\label{SAMDX}.
\end{equation}
The time-averaged linear and angular momenta can be obtained by
integrating over the whole $x-y$ plane~\cite{Jackson1999}
\begin{equation}
J_{a}=\int\int j_{a}\text{d}x \text{d}y\label{OAM},
\end{equation}
\begin{equation}
P_{a}=\int\int p_{a}\text{d}x \text{d}y\label{SAM}.
\end{equation}
The transverse orbital angular momentum would appear if the
spin-to-orbital angular momentum conversion occurs in the process of
reflection and refraction. It should be mentioned that the
transverse orbital angular momentum can lead to a transverse shift
of field centroid even in free space~\cite{Luo2010a}. In the plane
$z_a=0$, after substituting Eqs~(\ref{OAMDX}) and (\ref{SAMDX}) into
Eqs~(\ref{OAM}) and (\ref{SAM}) we obtain
\begin{equation}
\Delta
y_{a}=\frac{J_{ax}^S}{P_{az}^S}=\frac{J_{ax}^O}{P_{az}^O}\label{SAMM}.
\end{equation}
Figure~\ref{Fig5} plots the transverse spin and orbital momentum
currents for incident and transmitted fields. It is clearly shows
that the transverse orbital angular momentum plays a dominant role
in the SHE of light. Compare Fig~\ref{Fig5}(b) with (d), however, it
is not very clear why the transmitted field exhibits the SHE of
light. Thus, it is necessary to determine the transverse orbital
angular momentum. From Eq.~(\ref{OAM}), we can obtain $j_{ix}^{O}=0$
and $J_{tx}^{O}\neq0$. The presence of transverse orbital angular
momentum $J^O_{tx}$ provide a direct evidence for the
spin-to-orbital angular momentum conversion. In the SHE of light
associated with refraction~\cite{Hosten2008},
reflection~\cite{Qin2009}, and tunneling~\cite{Luo2010b}, the
transverse spatial shifts are sensitive to the incident polarization
angles.

We proceed to explore the role of refractive index gradient in
spin-to-orbital angular momentum conversion. The $z$ component of
total angular momentum $J_{az}$ for the $a$th beam can be
represented as a sum of the extrinsic orbital angular momentum
$J_{az}^O$ and the intrinsic spin angular momentum $J_{az}^S$, i.e.,
$J_{az}=J_{az}^O+J_{az}^S$~\cite{Bliokh2009}. The $z$ component of
the orbital angular momenta are given by $J_{iz}=0$, $J_{rz}=-\Delta
y_r k_r \sin\theta_r$, and $J_{tz}=-\Delta y_t k_t \sin\theta_t$.
for incident, reflected, and transmitted wave packets, respectively.
In the laboratory Cartesian frame ($x,y,z$), the $z$ component of
the orbital angular momenta are given by
\begin{equation}
J_{iz\pm}^O=0\label{JIZO},
\end{equation}
\begin{equation}
J_{rz\pm}^O=\pm\frac{(r_p^2+r_s^2-2r_p r_s)\cos\theta_i}{ r_p^2 +
r_s^2}\label{JRZO},
\end{equation}
\begin{equation}
J_{zt\pm}^O=\pm\frac{(t_p^2+t_s^2-2\eta t_pt_s)\cos\theta_i}{ t_p^2
+t_s^2}\label{JTZO}.
\end{equation}
The $z_a$ components of spin angular momentum for the $a$th beam is
respectively described by
\begin{equation}
J_{iz\pm}^S=\pm\cos\theta_i\label{JIZS},
\end{equation}
\begin{equation}
J_{rz\pm}^S=\pm\frac{2 r_p r_s}{r_p^2 +
r_s^2}\cos\theta_r\label{JRZS},
\end{equation}
\begin{equation}
J_{tz\pm}^S=\pm\frac{2 t_p t_s}{t_p^2 +
t_s^2}\cos\theta_t\label{JTZS}.
\end{equation}
From Eqs.~(\ref{JIZO})-(\ref{JTZO}) and
Eqs.~(\ref{JIZS})-(\ref{JTZS}), we find that the angular momenta
fulfill the conservation law:
\begin{equation}
Q_r J_{rz\pm}+ Q_t J_{tz\pm}=J_{iz\pm}.\label{totalam}
\end{equation}
Here, $Q_r=(r_p^2+r_s^2)/2$ and $Q_t=n\eta(t_p^2+t_s^2)/2$ are the
energy reflection and energy transmission coefficients,
respectively. Note that the transverse angular shifts, which
governed by the linear momentum conservation law, are not discussed
in this paper.

\begin{figure}
\includegraphics[width=8cm]{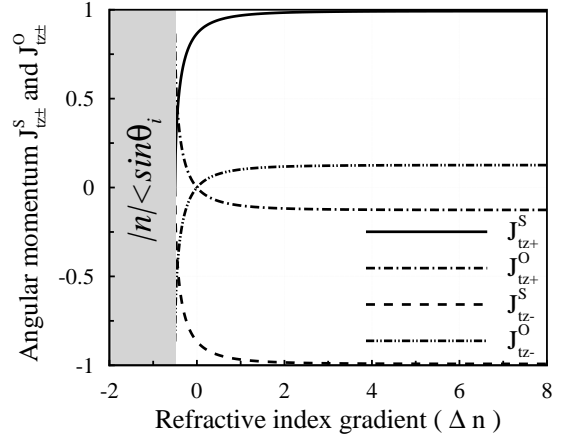}
\caption{\label{Fig6} The refractive index gradient $\Delta{n}$
induces the spin-to-orbital angular momentum conversion. The
incident wave packets are chosen as left circular polarization and
right circular polarization. Other parameters are chosen to be the
same as in Fig.~\ref{Fig2}.}
\end{figure}

To obtain a clear physical picture of the SHE of light, the
spin-to-orbital angular momentum conversion induced by refractive
index gradient is depicted in Fig.~\ref{Fig6}. For the left
circularly polarized component, the spin angular momentum
$J^S_{tz+}$ monotonously increase while the orbital angular momentum
$J^O_{tz+}$ monotonously decreases with the increase of the
refractive index gradient $\Delta n$. For the right circular
polarization, both the spin angular momentum $J^S_{tz-}$ and the
orbital one $J^O_{tz-}$ present oppositive features. When the
refractive index gradient continues increasing, the spin-to-orbital
angular momentum conversion appears to reach a saturation value:
\begin{equation}
J^O_{tz\pm}=\mp\frac{\cos\theta_i\sin^2\theta_i}
{1+\cos^2\theta_i}\label{JOTZM}.
\end{equation}
This gives a good explanation of the intriguing phenomena why the
transverse spatial shifts tends to saturation values in the
Fig.~\ref{Fig3}. The metamaterial with $n=-1$ is a good candidate to
eliminate the SHE of light in the refraction, since the
spin-to-orbital angular momentum conversion is impressed completely.
The spin-to-orbital angular momentum conversion can be enhanced in
the region of low refractive index gradient as shown in
Fig.~\ref{Fig6}. As a result, the transverse spatial shifts are also
be amplified in this region. However, when the refractive index
$|n|<\sin\theta_i$ the wave pack is totally reflected.  By properly
facilitating the spin-to-orbital angular momentum conversion the SHE
may be enhanced dramatically.

We now give a very simple way to understand how the refractive index
gradient enhance the spin-to-orbital angular momentum conversion in
the SHE of light. We attempt to perform analyses on the $z$
component of the total angular momentum for each of individual
photons, i.e., $J_{iz}=J_{tz}$~\cite{Onoda2006,Nasalski2006}. The
total angular momentum conservation law for single photon is given
by
\begin{equation}
J^O_{tz\pm}\pm\cos\theta_{t}=\pm\cos\theta_i,
\end{equation}
where $J^O_{tz\pm}=-\Delta y_{t\pm} k_t \sin\theta_t$. When the
photons penetrate from air into a low-refractive-index medium
($\Delta n<0$), the incident angle is less than the transmitted
angle $\theta_i<\theta_t$. For the left circularly polarized
photons, the $z_t$ component of spin angular momentum
$+\cos\theta_{t}$ decreases after entering the medium. Because of
the conservation law, the total angular momentum must remain
unchanged. To conserve the total angular momentum, the photons must
move to the direction $-y$ ($\Delta y_{t+}<0$) and thus generate a
positive orbital angular momentum ($J_{tz+}^O>0$). For the right
circularly polarized photons, the $z$ component of spin angular
momentum $-\cos\theta_t$ increases. In this case, the photons must
move to the direction $+y$ ($\Delta y_{t-}>0$) and induce a negative
orbital angular momentum ($J_{tz-}^O<0$). When the photons enter
into a high-refractive-index medium ($\Delta n>0$), the incident
angle is larger than the transmitted angle $\theta_i>\theta_t$. As a
result, the orbital angular momentum reverses its direction.

\section{Conclusions}
In conclusion, we have identified the role of spin-to-orbital
angular momentum conversion in SHE of light. We have demonstrated
that the refractive index gradient can enhance or suppress the
spin-to-orbital angular momentum conversion, and thus can control
the SHE of light. The recent advent of metamaterial whose refractive
index can be tailored arbitrarily seems to be an available candidate
to amplify or eliminate the SHE of light. However, the
spin-to-orbital angular momentum conversion in a ultra-large
refractive index gradient is limited by a saturation value.
Fortunately, the SHE of light can be dramatically amplified by
plasmonic
nanostructure~\cite{Gorodetski2008,Gorodetski2009,Vuong2010,Herrera2010}.
In addition, the SHE of light can also be noticeably enhanced when
the beam carries orbital angular
momentum~\cite{Bliokh2006b,Okuda2006,Fadeyeva2009,Merano2010}.
Hence, the exploration of spin-to-orbital angular momentum
conversion in these systems would be very interesting. The
transverse spatial shifts governed by the spin-to-orbital angular
momentum conversion, give us a clear physical picture to clarify the
role of refractive index gradient in the SHE of light. These
findings provide a pathway for modulating the SHE of light, and
thereby open the possibility for developing new nano-photonic
devices. Because of the close similarity in optical
physics~\cite{Hosten2008,Bliokh2008}, condensed
matter~\cite{Murakami2003,Sinova2004,Wunderlich2005}, and
high-energy physics~\cite{Berard2006,Gosselin2007}, by properly
facilitating the spin-to-orbital angular momentum conversion, the
SHE may be enhanced dramatically in these physical systems.

\begin{acknowledgements}
One of authors (H.L.) thanks Dr. W. L\"{o}ffler for helpful
discussions. This research was partially supported by the National
Natural Science Foundation of China (Grants Nos. 61025024, 11074068,
and 10904036).
\end{acknowledgements}

\end{document}